\documentclass[aps,prb,twocolumn,groupedaddress,showpacs]{revtex4}
\usepackage{psfig}
\usepackage{epsfig}

\newcommand{\beq}{\begin{equation}}
\newcommand{\eeq}{\end{equation}}
\newcommand{\bea}{\begin{eqnarray}}
\newcommand{\eea}{\end{eqnarray}}
\newcommand{\nn}{\nonumber}

\newcommand{\eps}{\epsilon}
\newcommand{\veps}{\varepsilon}
\newcommand{\al}{\alpha}
\newcommand{\s}{\sigma}
\newcommand{\lam}{\lambda}
\newcommand{\de}{\delta}
\newcommand{\D}{\Delta}

\newcommand{\Ga}{\Gamma}
\newcommand{\ga}{\gamma}

\newcommand{\ua}{\uparrow}
\newcommand{\da}{\downarrow}

\newcommand{\lra}{\longrightarrow}

\begin{document}
\bibliographystyle{apsrev}
\title{Kondo screening cloud in a double-quantum dot system}

\author{Pascal Simon}
\email{pascal.simon@grenoble.cnrs.fr}
\affiliation{ Laboratoire de Physique et Mod\'elisation des Milieux  Condens\'es\\
CNRS and University Joseph Fourier, BP 166, 25 av. des martyrs, 38042 Grenoble, France}
\date{\today} 
\begin{abstract}
We analyze the transport properties of two artificial magnetic impurities coupled together
via a tunable RKKY interaction mediated by conduction electrons of
 a finite size one dimensional wire.
We show that the sign of the RKKY interaction can be controlled by gating the wire. We investigate
 the interplay between finite size effects and  RKKY interaction and found that the two artificial impurities start to interact each-other as soon as the Kondo screening cloud length becomes larger 
that the length
of the wire. This should allow to give a lower experimental estimate of the Kondo screening cloud length.
\end{abstract}

\maketitle

\section{Introduction}

One of the most remarkable triumphs of recent progress in nanoelectronics 
has been the observation of the Kondo effect in a single semi-conductor 
quantum dot. \cite{dot,Cronenwett,Wiel}   When the number of electrons in 
the dot is odd, it can behave as an $S=1/2$ magnetic impurity interacting via 
magnetic exchange with the conduction electrons. One of the main signatures
of the Kondo effect is that the conductance through the quantum dot reaches the named unitary
limit $2e^2/h$ at low enough temperature $T<T_K^0$ where $T_K^0$ is the 
Kondo temperature. In this temperature regime, the impurity spin is screened and 
forms a singlet with a conduction electron belonging to a very extended many-body
wave-function known as the Kondo screening cloud. 
The 
size of this screening cloud may be evaluated as $\xi_K^0\approx \hbar v_F/T_K^0$ where 
$v_F$ is the Fermi velocity. In a quantum dot, the typical Kondo temperature is of order
$1~K$ which leads to $\xi_K^0\approx 1$ micron in semiconducting heterostructures.
Finite size effects related to the final extent of this length scale have been predicted
recently in two different geometries: a quantum dot embedded in a ring threaded by a magnetic flux \cite{affleck01,Hu,sorensen04}
and a quantum dot embedded between two finite size wires connected to  external leads.\cite{simon02,simon03,balseiro}
In the former geometry, it was shown that the persistent 
current induced by a magnetic flux is particularly sensitive to screening 
cloud effects and is drastically reduced when the circumference 
of the ring becomes smaller than $\xi_K^0$. \cite{affleck01} In the latter geometry,  signature of the finite size extension  of the Kondo cloud was found in the temperature dependence of the 
 conductance through the whole system (wires+dot).\cite{simon02,simon03,balseiro}

The previous results were obtained by considering a single impurity confined in some mesoscopic finite size environment. Similar effects have also been predicted in related geometries.\cite{thimm,balseiro02}

Nevertheless, a more appropriate and certainly relevant question for bulk materials 
would be what is the lower bound on the
average impurity separation necessary to apply the single impurity theory ?
As described in [\onlinecite{barzykin}], there are at least three different ways of estimating such a length scale:
i) One can first simply require that the minimum inter-impurity distance to be larger than $\xi_K^0$.
ii) One may require that the density of electron states within an energy $T_K^0$ of the Fermi surface be at least as large as $n_i$, the impurity concentration (this constitutes the so called Nozi\`eres exhaustion principle).
iii) Following the mean field treatment of the Kondo lattice by Doniach,\cite{doniach} one may also require that the energy scale related to the average RKKY interaction to be small compared to $T_K^0$. 

In one dimension (1D), definitions (i) and (ii) provide $R\sim \xi_K^0$ for the minimum distance at which
impurities can be treated independently whereas definition (iii) provides $R\sim (\rho^0 J)^2 \xi_K^0$ where
$J$ is the Kondo coupling between the impurity spin and the conduction electrons and $\rho^0$ the density of states. Definition (iii) gives therefore a lower estimate for the inter-impurity distance to account for multi-impurity effects. 

Recent progress in nanolithography allows to design complex geometries able to 
mimic the behavior of a 
many impurity problem. In particular, Craig et al. \cite{craig} have recently manufactured a system in which two Kondo quantum dots are connected to a a common open conducting region. Here both the RKKY and Kondo interactions compete, thus providing an experimental realization of the two-impurity Kondo problem. 

\begin{figure}
\epsfig{figure=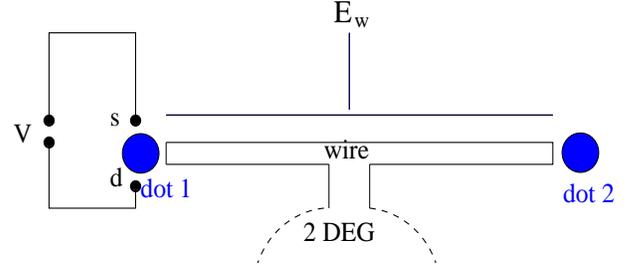, width=8cm,height=3.5cm}
\caption{Schematic representation of the device under study: the two quantum dots are connected to a finite size open wire.
The Fermi level in the wire is controlled by a gate voltage $E_W$. }
\label{Fig:scheme}
\end{figure}
In this paper, we propose to use a rather similar geometry to estimate the Kondo screening cloud size.
We consider the following device: two quantum dots connected to a quantum wire
of length $2l$. We assume that the quantum dots can be both tuned via independent plunger gate voltages in the Kondo regime, and that
the finite size wire (FSW) can be gated. We also assume the FSW  to be weakly connected to a two dimensional reservoir. 
The whole geometry 
is depicted in the Figure \ref{Fig:scheme}. The setup we want to study is
therefore related to a two impurity Kondo problem in a confining geometry, a question recently addressed by Galkin et al.\cite{galkin} using slave boson mean field theory. Galkin et al. have shown that the low energy behavior of this system
strongly depends on the position of the Fermi energy
as for the single impurity problem in a confined geometry.\cite{galkin}

We intend to show that when $2l\gg\xi_K^0$, the impurities behave almost independently whereas for $2l\lesssim \xi_K^0$, $2-$impurity effect takes place. The analysis
of transport through one of the quantum dot allows to discriminate between a single impurity behavior from
a 2-impurity one and may therefore provides an estimate of $\xi_K^0$.
The plan of the paper is the following: in section 2, we present our model Hamiltonian
and derive how the finite size effects (FSE) renormalize 
both the Kondo temperature and the RKKY interaction.
In section 3, we show FSE affect quantitatively the transport properties through the quantum dot. Finally in section 4, we give a brief conclusion and 
discuss the generality of our results.

\section{Finite size effects and RKKY interactions}
In such a geometry, the two artificial impurities are connected to both extremities of a quantum gated wire.
Since the wire is also weakly connected to a two dimensional reservoir, its spectrum is continuous.
In such a confined geometry, FSE are to play an important role, first because they 
 affect
the renormalization of the Kondo couplings of each impurities as was shown in [\onlinecite{simon02}] but also the RKKY interaction. To take both effects into account, we choose to use a simple tight-binding model which
should capture the main physical ingredients of this system.

\subsection{Model Hamiltonian}

In order to treat this problem, we may write the following simple  $1D$ tight-binding Hamiltonian:
\beq 
H=H_D+H_W+H_R+H_{tun,D}+H_{tun,R}
\eeq
with
\bea
H_D&=& \sum\limits_{i=1,2;\s} \eps_{d,i} d^\dag_{i,\s}d_{i,\s} + U_i n_{d,i,\ua}n_{d,i,\da}~,\label{hd}\\
H_W&=&-t\sum\limits_{j=1;\s}^{2l} (c_{j,\s}^\dag c_{j+1,\s} +h.c.) 
   + E_W\sum\limits_{j=1;\s}^{2l} n_{c,j,\s}~,\label{hw}\\
H_R&=&-t\sum\limits_{l=1}^{\infty}\sum\limits_{\al=1,2;\s} (a_{l,\al,\s}^\dag a_{l+1,\al,\s} +h.c.)~, \label{hr}\\
H_{tun,D}&=&-t'\sum\limits_\s (c_{1,\s}^\dag d_{1,\s}+c_{2l,\s}^\dag d_{2,\s}+h.c.)~,\label{htund}\\
H_{tun,R}&=&-t_W\sum\limits_\s(c^\dag_{l,\s}a_{1,1,\s}+(c^\dag_{l+1,\s}a_{1,2,\s}+h.c.)\label{htunr},
\eea
where $d^\dag_{i,\s}$ creates an electron with spin $\s$ in dot $i$, $c_{j,\s}^\dag$ creates an electron in the wire at site $j$ with spin $\s$ and finally $a_{l,\al,\s}^\dag$ creates an electron in the reservoir at site $l$, in the channel $\al$ with spin $\s$.
In Eq. (\ref{hd}), the two quantum dots are described by two Anderson impurity models where
$\eps_{d,i},U_i$ are respectively the energy level and the Coulomb repulsion energy in dot $i$.
Eq. (\ref{hw}) describes a $1D$ non interacting finite size wire (FSW) of length $2l$.
 In Eq. (\ref{hr}), we described the reservoir that coupled to the FSW by an infinite
 two channels wire. This assumption is for further technical convenience, the only ingredient required being 
that the FSW is tunnel coupled to a continuum (see Ref. [\onlinecite{simon03}] for details).
The last two equations (\ref{htund}) and (\ref{htunr})  describe tunneling of electrons respectively between the dots and the FSW and between the reservoir and the FSW.
In the following we also 
 assume a symmetric geometry {\it i.e} we suppose $\eps_{d,1}=\eps_{d,2}=\eps_d$ and $U_1=U_2=U$. In fact, we only require in the sequel
that  the Kondo temperatures associated with each dot are comparable {\it i.e.} $T_K^0=T_{K,1}^0\sim T_{K,2}^0$. 
We have also assumed that the FSW is connected to a gate controlled
by $E_W$. In order to study transport properties, we  need to weakly connect
at least one dot to extra leads (that are represented with dashed lines in the figure  \ref{Fig:scheme}).
Such way of probing the quantum dot transport properties is more in the spirit of the STM type measurements since it directly
probes the local dot density of states. 

We assume that the two dots can be cast in the Kondo regime. Such situation has been recently realized experimentally in Ref. [\onlinecite{craig}]. We can perform a Schrieffer Wolff transformation in order to write $H_D+H_{tun,D}$ in the following form:
\beq \label{hkondo}
H_D+H_{tun,D}\lra J( \vec S_1\cdot \s_1 +\vec S_2\cdot\s_{2l})~,
\eeq 
with $J=2t'^2\left(\frac{1}{-\eps_d}+\frac{1}{U+\eps_d}\right)$ and 
$\vec \s_{1/2l}=c^\dag_{1/2l}\frac{\vec \tau}{2}c_{1/2l}$ is the local spin density at sites $1/2l$ ($\vec\tau$ are the Pauli matrices). We have  neglected potential scattering terms which are not playing an important role
in what follows.

Before taking into account both the Kondo and RKKY interactions, we need to evaluate these two interactions
in a confined geometry. 

\subsection{Kondo temperatures}
Let us first estimate the Kondo temperature of a {\it single} artificial impurity
assuming that one quantum dot is disconnected of the system. This can be achieved by pinching the wire-dot tunnel junction off. The remaining dot is coupled to the FSW and its Kondo temperature can be estimated following Ref. [\onlinecite{simon03}].
In order to calculate $T_K$, we first diagonalize the Hamiltonian 
at $J=0$, i.e. we diagonalize 
$H_0\equiv H_W+H_{R}+H_{tun,R}$.
If $t_{W}=0$, then the wave-functions 
and eigenvalues of the wire  are:
\begin{eqnarray} \label{eigen}
\psi_(j)&=&(1/\sqrt{2l+1})\sin( k_{w,n}j )~, \nn \\
k_{w,n}&=&\pi n/(2l+1);~~1\leq n\leq 2l ~,\\ 
\eps(k_{w,n})&=& -2t\cos k_{w,n}+E_W. \nn\end{eqnarray}  

For non-zero $t_{W}$, the spectrum of $H_0$  becomes continuous.
In order to study how the Kondo interaction renormalizes, 
we express $c_{1},c_{2l}$ in terms of the  eigenstates, 
$c_\epsilon$ of $H_0$:
\beq\label{deff}
c_{j} = \int_{-2t}^{2t} d\epsilon \left(f_j^e(\epsilon)c_{e,\epsilon}+
f_j^o(\epsilon)c_{o,\epsilon}\right),
\eeq
with $j=1,2l$ here and the indices $e,o$ are for even or odd components under parity symmetry under the middle of the line. Indeed $H_0$ can be decomposed as $H_0^e+H_0^o$ which are obtained by a folding procedure around a vertical axis located at the point $j=l+1/2$.
We have normalized the operators $c_{e/o,\eps}$ such that $\{c_{e/o,\eps}^\dagger,c_{e/o,\eps'}\}=\delta(\eps-\eps')$ and $\{c_{e/o,\eps}^\dagger,c_{o/e,\eps'}\}=0$.
The local density of states seen by dot 1 is defined by 
\beq
\rho_1(\eps)=|f_1^e(\eps)|^2+|f_1^o(\eps)|^2=\rho_1^e(\eps)+\rho_1^o(\eps)\eeq
and is normalized according to $\int_{-2t}^{2t} d\epsilon \rho_1(\eps)=1.$
 In this basis the Kondo Hamiltonian (\ref{hkondo}) can be written as $H=H_0+H_K$
with
\beq
H_0=\int d\eps \eps\left(c_{e,\eps}^\dag c_{e,\eps}+c_{o,\eps}^\dag c_{o,\eps}\right),
\eeq
and
\bea\label{hk}
H_K&=&=\int\int d\eps ~d\eps'\left\{ [f_1^e(\eps)f_1^e(\eps') c_{e,\eps}^\dagger\vec \s c_{e,\eps'}
\right.\nn\\
&&+f_1^o(\eps) f_1^o(\eps') c_{o,\eps}^\dagger\vec \s c_{o,\eps'}]\cdot(\vec S_1+\vec S_2)\nn\\
&&+f_1^e(\eps) f_1^o(\eps')\times \\
&&\left.[c_{e,\eps}^\dagger\vec \s c_{o,\eps'}+c_{o,\eps}^\dagger\vec \s c_{e,\eps'}]
\cdot (\vec S_1-\vec S_2)\right\}.\nn
\eea
To obtain this equation, we have used the fact that $f_{2l}^e(\eps)=f_1^e(\eps)$ and $f_{2l}^o(\eps)=-f_1^o(\eps)$.

\begin{figure}
\epsfig{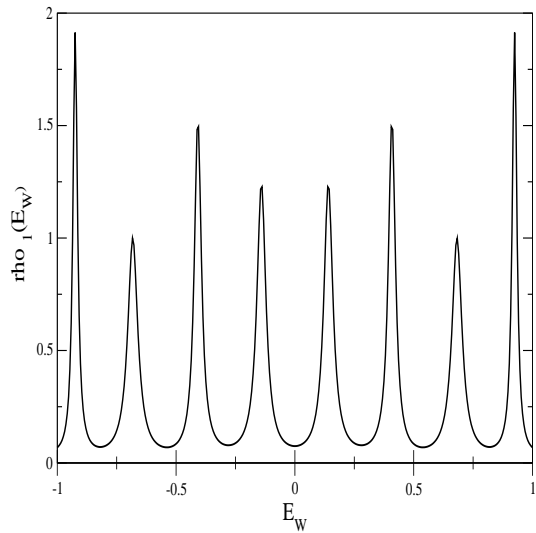}
\caption{Example of the local density of states seen by the dot 1. We took $2l=22$ and $\ga_W^2=0.25$}
\label{Fig:ldos}
\end{figure}

The local density of states $\rho_1(\eps)$ can be computed exactly for our tight binding model following Ref. [\onlinecite{simon02}] and reads:
\beq
\rho_1^{even/odd}(k)=\frac{1}{\pi t}\frac{\ga_W^2\sin^2 k_w\sin k}{D_{e/o}^2}~,
\eeq
with
\bea
D_{e/o}^2 &=&\left[\cos kl(\sin k_w(l+1)\mp \sin k_wl)\right. \nn \\
&-&  \left.\gamma_W^2\cos k(l+1)\sin(k_w l)\right]^2\nn\\
&+& \left[\sin kl( \sin k_w(l+1)\mp \sin k_wl)\right. \nn \\
&-& \left. \gamma_W^2 \sin k(l+1)\sin(k_w l)\right]^2\nn.
\eea
Note that we  have defined $\gamma_W=t_W/t$ and $k_w$ is related to $k$ by 
\beq \label{defkw}
-2t\cos k=-2t\cos k_w +E_W.
\eeq
If $\ga_W=0$, the even LDOS $\rho_1^e$ is simply composed of $l$ Dirac peaks/levels separated
by the level spacing $2\Delta\sim \hbar v_F/l$. $\rho_1^o$ has a similar structure but is shifted by
$\Delta$. When $\ga_W\neq 0$, these $2l$ finite size levels  in the wire acquire  
a finite width and their position is slightly shifted because of the coupling to the continuum spectrum. $\rho_1^{e/o}$ become now continuous. An example of such a local density of states
has been plotted for $l=11$ and $\ga_W=0.5$ in Fig. \ref{Fig:ldos}.
Note that even for $\ga_W=1$, i.e. a perfect coupling between the source and the leads,
we get a succession of peaks of broader widths. We have modeled the reservoir by a two channel infinite wire 
in order that both  even and odd sectors couple  to the continuum.
This density profile is quite general and does not depend on our particular model Hamiltonian.
It occurs as soon as we couple a finite size system to a continuum.

New energy scales emerge in the problem, the level spacing $\D$, which needs to be compared
to the bare Kondo temperature $T_K^0$ of the dots and the peak width $\delta$. By bare Kondo temperature $T_K^0$, we mean
the Kondo temperature of a single independent dot in the Kondo regime i.e. the Kondo temperature
obtained in the limit $l\to \infty$.
When $T_K^0\sim \D$, FSE start taking place. The genuine
Kondo temperature of a single dot $T_K$ starts to depend on the fine structure of the local density of states and can be in general very different from $T_K^0$. 
In this situation, the Kondo temperature $T_K$ depends on the wire gate voltage $E_W$
being set on a resonance (which gives a Kondo temperature $T_K^R$) or off a resonance
 (which gives a Kondo temperature $T_K^{NR}$). This discussion has been largely detailed in Ref. [\onlinecite{simon02,simon03}]. In the limite $\gamma_W^2\ll 1$ one can evaluate:
\beq
T_K^{R}\approx \de\left( {T^0_K\over
D_0}\right)^{\gamma_{W}^2}=O(\de_n),
\eeq
 and 
\beq
T_K^{NR}\approx \Delta\left({T_K^0\over D_0}\right)^{1/2\gamma_{W}^2}.
\label{tkor}
\eeq  
For strong finite effects, one has $T_K^{NR}\ll T_K^R\sim O(\delta)$. Such dramatic dependence of the single impurity energy scale
with the gate voltage position clearly affects all transport and thermodynamic properties.\cite{simon02,simon03}

When both impurities are taken into account, the previous results can be extended starting from the hamiltonian in Eq. (\ref{hk}). If the RKKY interactions were switched off, both impurities would be now coherently screened in a two stage procedure. As above, FSE take place when $T_K^0\sim \Delta$. When this situation is met, the two energy scales associated to the two stage screening process depend strongly on the even/odd local densities of states being on or off resonances. This situation is similar to the one occuring for a ferromagnetic RKKY interaction and will be discussed in section \ref{discussion}.

\subsection{RKKY interaction}

In the previous subsection, we have focussed on the Kondo temperature of a single dot ignoring
their mutual interactions. Nevertheless, 
the two dots  also interact through the RKKY
interaction. The RKKY interaction reads: $H_{RKKY}=I \vec S_1\cdot\vec S_2$.
This interaction is automatically generated at second order in the Kondo coupling $J$ and corresponds to the coherent process where an electron participating in the screening of the artificial first impurity  can also participate in the screening of the second impurity. Remmenber that  a more dynamical type of screening is involved
here rather than the one occuring for charge impurities in a Fermi liquid.

\subsubsection{Usual situation: $T_K^0\gg \D$}

Let us first consider the simple case of a very long wire connecting the two dots without the reservoir ({\it i.e.} $\ga_W=0$).
At lowest order in $J$, the RKKY interaction reads
\bea
I=&&J^2 \left(\frac{2}{2l}\right)^2\times\\
&&\sum\limits_{\eps(k_w)<\eps_F;\eps(k'_w)>\eps_F}
\frac{\sin(k_w) \sin (k'_w)\sin(2k_wl)\sin(2k'_wl)}{\eps(k_w)-\eps(k'_w)},\nn
\eea
where the $\eps(k_w)$ have been defined in Eq. (\ref{eigen}) and $\eps_F$ is the Fermi energy.
For $l\gg 1$ and $\eps_F\sim 0$, this expression simplifies 
and reads:\cite{sorensen96}
\beq
I= \frac{J^2\cos(2k_F(2l))}{4\pi t(2l)}=\frac{\pi \lam^2 D\cos(2k_F(2l))}{8(2l)},\eeq
where $\lam=\rho_0 J=J\sin(k_F)/(\pi t)$ is the dimensionless Kondo coupling and $D=2t$ is the band width.
This is the usual from of the RKKY interaction in $1D$. The above interaction is always positive
at exactly half filling because we take $2l$ sites. Had we we taken an odd number of sites, we would get an opposite sign i.e a ferromagnetic interaction.

One may argue that the impurity can no longer be regarded as independent when $I\sim T_K^0$.
This result has been obtained for example by applying the slave boson mean field theory
approximation for the two impurity problem by Jones {\it et al.}. \cite{jones89}
From this argument, we have seen in the introduction that at fixed $\xi_K^0$, the dots need
to be brought rather close (compared to $\xi_K^0$) to start feeling the RKKY interaction effects.
However, we have seen that screening cloud FSE occur already when $\xi_K^0\sim 2l$ with a strong renormalization of $T_K$. This may suggest that FSE occur first when approaching
the two impurities. Nevertheless, this argument is too naive since FSE may also affect the RKKY interaction as we will see. A more refined analysis is clearly demanded
when $T_K^0\lesssim  \D$.

\subsubsection{Unusual situation: $T_K^0\ll \D$}
 
In this regime, one have already seen how the Kondo temperature is renormalized depending
on the density of states being on resonance or off resonance.
In our general case of a finite size wire connected to a source lead, the 
RKKY interaction can be written at lowest order in $J$ as:
\bea
H_{RKKY}&&=I\vec S_1\vec S_2=\\
&&J^2 \vec S_1\vec S_2\int\limits_{-D}^{\eps_F}d\veps\int\limits_{\eps_F}^{D}d\eps'
\frac{ f_1(\eps)f_{2l}(\eps)f_1(\eps')f_{2l}(\eps')}{\eps-\eps'}\nn
\eea
where the $f_j(\eps)$ have been defined in Eq.( \ref{deff}).
In the limit $\ga_W\ll 1$, the LDOS they can be well approximated as follow:
\begin{widetext}
\beq
\pi \rho_j(\eps)=\pi f_j^2(\eps)\approx \frac{2}{2l+1}\sum_n \sin^2(k_{w,n} j) \frac{\delta_n}{(\eps-\eps_n)^2+\delta_n^2},
\eeq
and in a similar way
\beq
\pi f_1(\eps)f_{2l}(\eps)\approx  \frac{2}{2l+1}\sum_n \sin(k_{w,n})\sin(2l k_{w,n}) \frac{\delta_n}{(\eps-\eps_n)^2+\delta_n^2}.\eeq

In these expressions, we have defined $\eps_n=\eps(k_{w,n})$ ($(k_{w,n}$ was introduced in  Eq. (\ref{eigen})) and $\delta_n\approx \ga_W^2 t\sin^2k_{w,n} \sin k/l$ corresponds to the peak width.
This faithful approximation allows us to evaluate the RKKY interaction when finite size
effects do occur:

\beq \label{rkkyapprox}
I\approx \left(\frac{J}{\pi}\right)^2\left(\frac{2}{2l+1}\right)^2\sum\limits_{n,n'}\int\limits_{-D}^{\eps_F}  d\eps \int\limits_{\eps_F}^D d\eps'
sin^2(k_n)sin^2(k_{n'}) (-1)^{n+n'}\frac{\de_n\de_{n'}}{(\eps-\eps')((\eps-\eps_n)^2+\de_n^2)
((\eps'-\eps_{n'})^2+\de_{n'}^2)}.
\eeq
\end{widetext}

\begin{figure}
\epsfig{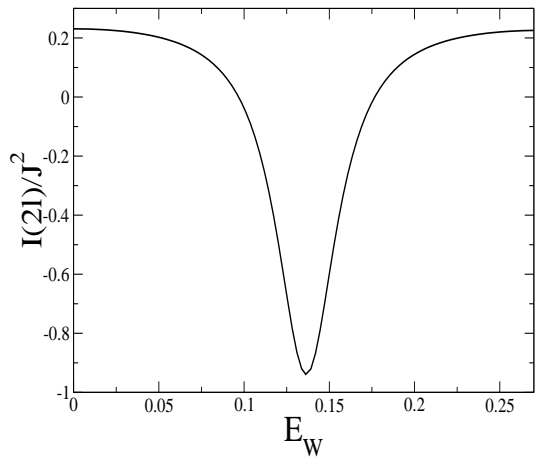}
\caption{RKKY interaction as a function of $E_W$. We took the same parameter as in Fig. \ref{Fig:ldos}: 
 $2l=22$ and $\ga_W^2=0.25$. When $E_W$ is tuned on a resonance, the RKKY interaction becomes positive and gets enhanced by a factor  $\sim 1/\ga_W^2$ as predicted by Eq. (\ref{eq:ir}).}
\label{Fig:rkky}
\end{figure}

The terms that contribute mostly in the above summation are the ones where $\eps_n$ are in the vicinity of $\eps_F$. 
We need to distinguish two different
situations:

(i) Suppose that $E_W$ is tuned such that $\eps_F$ lies between two resonances peaks
of the local density of states. The integral is therefore mainly dominated by these two consecutive resonance peaks that implies that $n+n'$ is odd. The whole integral is therefore
positive meaning the RKKY interaction is antiferromagnetic.

(ii) Suppose that  $E_W$ is tuned such that $\eps_F$ is on a resonance.
The integral is mainly dominated by this resonance peak and therefore $n=n'=n_F$.
The whole integral is therefore negative meaning the RKKY interaction is ferromagnetic.

We can therefore conclude that FSE also plays a crucial
role by fixing the sign of the RKKY interaction as already  noted in [\onlinecite{galkin}]. When the LDOS is tuned off resonance, the RKKY interaction
is always antiferromagnetic, when the  LDOS  is tuned on resonance, the RKKY interaction
is always ferromagnetic. This strong sign dependence opens the way for a control of the RKKY interaction.

Furthermore, these integrals can be estimated (in the limit $\gamma_W\ll 1$) using the previous approximations:
\bea
I^{NR}&\approx& J^2 \frac{2 c_1\sin^4(k_F)}{(2l)^2\D}\approx J^2\frac{ c_1\sin^2k_F\rho_0}{ 2l},\\
I^R&\approx& -J^2\frac{ c_2\sin(k_F)}{2l\pi \ga_W^2t}\approx 
 -J^2\frac{ c_2\rho_0}{2l\ga_W^2},\label{eq:ir}
\eea
with $\rho_0=\sin(k_F)/(\pi t)$ and $c_1,c_2$ two numerical factors of order 1. 

We see that the RKKY interaction $I^R$ is actually enhanced when the LDOS is tuned on resonance
($\ga_W^2\ll 1$) whereas $I^{NR}\approx I$ remains almost unchanged. A rather similar behavior
occurs for magnetic impurities in quantum corrals.\cite{correa} In order to check these predictions, we 
have calculated numerically the integral given by Eq. (\ref{rkkyapprox}) and plotted
the RKKY
interaction $I$ as a function of $E_W$ in Fig. \ref{Fig:rkky} when $E_W$ goes through a resonance.
It confirms first that the RKKY interaction can be varied continuously from positive to negative value and
second that the RKKY interaction becomes ferromagnetic and considerably
enhanced when $E_W$ is nearby a resonance.

\subsection{Discussion}\label{discussion}

Let us discuss and compare the RKKY interaction with the Kondo interaction.
We assume that $J$ and $T_K^0\sim D\exp{-1/(\rho^0J)}$ are fixed and the distance between the dots can be varied.
When $T_K^0\gg \D$ i.e $\xi_K^0\ll (2l)$, we are in the usual situation where the impurities can be regarded
as almost independent (the overlap between both screening clouds is $O(2l/\xi_K^0$).\cite{barzykin}
The RKKY interaction becomes important when the distance $l$ between the dots satisfies $l\sim (\rho^0J)^2\xi_K^0$,
a condition that is clearly not satisfied in this limit.

Suppose now that we bring closer together the quantum dots such that $\xi_K^0 \ll 2l$.
 In order to analyze this limit, we may integrate out high energy degrees of freedom
resulting from a change of $J(D)=J$ to $J_{eff}(\D)$ in the above expressions for the RKKY
interaction. \cite{simon02}
The two scales that must be compared are now $T_K$ and $I$.

i) When the LDOS is tuned off resonance, $T_K^{NR}$ is strongly reduced  compared
to $T_K^0$ whereas $I^{NR}\sim I$ increased as $1/2l$ when reducing $l$.
 We therefore expect a sharp crossover  
from an independent regime to a a regime where the dots form a singlet (remember that
$I>0$ off resonance). This cross-over is mainly due the strong renormalization of $T_K^0$ into $T_K^{NR}$ rather than
to the increase of $I$.
 
ii) When the LDOS is tuned on resonance then $T_K^R=O(\delta)$ with $\delta$ the peak width
whereas the RKKY interaction is enhanced by the decrease of $l$ but especially by the
extra factor $\ga_W^{-2}\gg 1$. As we decrease $l$, we also expect here a sharp crossover from a two almost independent
impurity regime to a two ferromagnetically coupled impurity  regime as soon as $T_K^R\lesssim I^R$.
When two impurities are coupled ferromagnetically, they form an effective spin $1$ impurity.
Conduction electrons try to screen this effective spin 1 impurity.
Following the work by Jayaprakash {\it et al.},\cite{jayaprakash} we can write the following low energy effective Hamiltonian obtained by integrating out high energy degrees of freedom
from $D$ to $\Ga<\D$ around the resonance, where $\Ga$ is the scale at which $T_K^R\lesssim I^R$.
\bea
H_{eff}&\approx& H_{el}+J_{eff}(\Ga) \vec S\cdot (\vec\s_1+\vec\s_{2l})\nn\\
&\approx &H_{el} +J_{eff}(\Ga)\vec S\cdot (\s_{1}^{even}+\s_{1}^{odd}),
\eea
where $\vec \s^{even/odd}_1 $ correspond to the even/odd part of the local spin density
of states coupled to the effective impurity. As we have already seen,
this decomposition is obtained by folding the initial model under the vertical axis
$j=l+1/2$. The density of states consists of even/odd peaks. Therefore, we expect at very low energy a {\it two-stage} kondo effect in this regime. Indeed, suppose we adjust
$E_W$ on a resonance, this resonance can be either odd or even under parity.
Let us for example assume this resonance to be  even and keep on integrating out high
energy degrees of freedom. When we will be close to the resonance $J_{eff}\rho_{1}^e$ will grow rapidly whereas $J_{eff}\rho_{1}^o$ will remain rather small.
In the limit $\ga_W\ll 1$ these two scales can be evaluated easily and correspond approximately to $T_K^R$ and  $(T_K^{NR})^2/\D<T_K^{NR}$. We note that the ratio between these two scales is exponential contrary to what was claimed in [\onlinecite{galkin}].
  
In the previous paragraph we compared $\xi_K^0$ and $2l$. It is of course not obvious to fine-tune $2l$ the wire length. On the other hand, we may find for each dot different Kondo valleys characterized by different Kondo
temperatures and therefore different Kondo screening lengths. A natural way to proceed experimentally
would be to select for each dot, at least two distinct Kondo valleys with rather different Kondo scales (ideally by one order of magnitude) in order
to probe both regimes $\xi_K^0=\xi_{K,1}^0\sim\xi_{K,2}^0\gg 2l$ and $\xi_K^0=\xi_{K,1}^0\sim\xi_{K,2}^0 \lesssim 2l$. Furthermore, in these Kondo valleys, the Kondo temperatures can also be varied using the dot
plunger gate voltages.

Finally, in the previous discussion, we assume $T=0$ and exhibit the main energy scales of the problem. An interesting and controversial issue is to obtain
quantitatively the full phase diagram (for example in the ($T,J$) plane) of 
the two impurity and especially Kondo lattice problem. In order to approach this non trivial issue, sophisticated techniques are required (see for example [\onlinecite{sun05}]).

\subsection{Non Fermi liquid fixed point}
In the  two impurities Kondo problem, the two Fermi liquid phases (the impurity singlet or the screened impurity triplet) may be separated at $T=0$ by a non 
Fermi liquid (NFL) fixed point depending on some realization of the particle-hole symmetry. \cite{jones,affleck,affleck95} One may wonder whether this  NFL fixed point can be reached in such a geometry.
We have seen that $2l\gg \xi_K^0$ is required in order for the RKKY interaction to take place.
In this regime, we need to distinguish between the LDOS being on or off a resonance.
When the LDOS is on a resonance, the RKKY interaction is ferromagnetic and no such NFL fixed point is found for this sign of the RKKY coupling. Suppose now the RKKY interaction is tuned in between two resonances
therefore of different parity. In our $1D$ geometry, this implies in particular
$f_1^e(-\eps)=f_1^o(\eps)$. When this condition is satisfied, it has been demonstrated by Affleck et al.,\cite{affleck95} that the Kondo Hamiltonian has indeed the wrong particle-hole symmetry and no nontrivial critical point  is expected to occur between the two FL phases. One can therefore conclude that such one dimensional like geometry is not suitable to reach the vicinity of the NFL fixed point of the two impurity Kondo problem. The only way to reach
such NFL fixed point would be to position the wire gate voltage on a resonance (therefore $I<0$) in order to have the right particle-hole symmetry  and to find an extra way to generate an additional antiferromagnetic coupling between the two impurities.

\section{Transport properties}
\subsection{Zero temperature}
The transport properties of the dots can be probed
individually  by attaching for example two very weakly coupled leads as depicted in Fig. 1.
The conductance is governed by the local density of the individual dots. Such a $3-$terminal measurement provides a spectroscopic analysis of the dot density of states much alike the STM probes the density of a magnetic impurity adsorbed on a surface.\cite{schiller}
From the analysis developed in the previous section, we infer that as soon as $\xi_K^0\ll (2l)$, the 
quantum dots are almost independent.
This corresponds to the usual situation
where the conductance reaches its maximum value. Extending the scattering approach developed by Ng and Lee \cite{ng} to several leads (see also [\onlinecite{simon03,mckenzie}], this maximum value reads:
\beq\label{gunitary}
G_U={2e^2\over h} \frac{4\Gamma_s\Gamma_d}{(\Gamma_d+\Gamma_s+\Gamma_W)^2}\approx {2e^2\over h} \frac{4\Gamma_s\Gamma_d}{\Gamma_W^2},
\eeq
where $\Gamma_W=\pi \rho^0 t'^2$ is the junction conductance between the dot and the wire and $\Gamma_{s/d}$ is the junction conductance between the dot and the source/drain lead.

The interesting situation occurs when $\xi_K^0\agt (2l)$.
We have seen that the sign and the amplitude of the RKKY interaction depends
on $E_W$ being adjusted to a resonance or off a resonance. 
This will have strong implications on transport properties.
In the former case (on resonance), the impurity is screened through a two stage procedure and the conductance 
can reach the maximum value given by Eq. (\ref{gunitary}). In the latter case, the impurities form a singlet
and the conductance is expected to be zero. Note that the conductance is in fact non zero for a generic quantum dot but instead dominated by potential scattering terms we neglected here that lead to elastic cotunneling processes. At $T=0$, the regime $\xi_K^0\gg 2l$ can be therefore characterized
by a variation of the dot conductance between $0$ and $G_U$ when tuning the wire gate voltage.  It is
worth looking at the differential conductance $dI/dV$ where V is the voltage bias between the two extra leads. A (narrow) 
peak at zero bias of width $T_K^R$ and height $G_U$ is expected in the former case whereas no peak is expected in the latter case. Furthermore, two extra small peaks are also expected in the differential conductance at $eV=\pm I$ corresponding to magnetic excitations. These features should allow to differentiate the single impurity behavior ($\xi_K^0\ll 2l$) from the 2-impurity behavior ($\xi_K^0\gg 2l$). The qualitative shape of these extra peaks has been analyzed in [\onlinecite{vavilov}].

\subsection{Finite temperature}
The previous results can be extended at finite temperature.
When $\xi_k^0\ll 2l$, the dots behave independently and the finite temperature conductance
of dot 1 simply reads
\beq
G_1=G_U f(T/T_K^0),
\eeq
where $f$ is a universal scaling function of the ratio $T/T_K$. Its asymptotic behavior reads:\cite{review}
\bea
f(x)&=&1-(\pi x)^2 ~~{\rm for}~~x\ll 1,\label{f-pert}\\
f(x)&=& \frac{3\pi^2/16}{\ln^2(x)}~~{\rm for}~~x\gg 1.\label{f-fl}
\eea  
On the other hand, in the limit $\xi_k^0\gg 2l$,  the conductance depends 
on the artificial impurities forming a singlet or a triplet state.
It is not obvious to read from the transport properties in which of these two states the impurities are,
as it was recently highlighted
by the experiment of Craig {\it et al.}. \cite{craig}
Indeed for $I<0$ (ferromagnetic RKKY interaction), the conductance takes the form 
\beq
G_1=G_U \frac{\pi^2/2}{\ln^2(T/T_K^R)}~~{\rm for}~~|I|\gg T\gg T_K^R,
\eeq  
and is therefore rather small compared to $G_U$.   The conductance for $I>0$ being dominated by
finite temperature cotunneling processes is rather small too. Therefore the linear conductance
does not really allow to discriminate in which states the impurities are in this temperature range.
However, as it was analyzed recently in [\onlinecite{vavilov,simon04}], transport spectroscopy in a finite in-plane magnetic field should enable to distinguish between both
impurity ground state (either singlet or triplet).
From an experimental point of view, it is therefore crucial to gate the wire since it may offer an
opportunity to go continuously from one regime to another way.

\section{Discussion of the results}
We have seen that when two quantum dots are brought closer together, they significantly deviate
from the single impurity behavior as soon as $\xi_K^0\gg 2l$. This deviation from the single impurity
behavior occur mainly because of the strong renormalization of the Kondo temperature but also from the renormalization of the RKKY interaction. When FSE play an important role, we also predict 
that the gating of the wire should allow
to control the sign of the RKKY interaction and therefore the wire gate voltage $E_W$ might be used
to control the ground state of our system (singlet/triplet).  By probing individually the spectral properties of each dot, one can differentiate the single impurity transport properties from the
2-impurities ones and therefore provide
a lower estimate of the Kondo screening length.

In order to obtain these results we made some approximations that we want to discuss here.
First we assume that the wire is thin enough such that we can restrict a single transverse channel.
If the wire contains $N$ different transverse channels, the results derived here can be extended following [\onlinecite{simon02}]. In particular, FSE start to play a role
when $T_K^0\lesssim \D/N$ i.e. for $\xi_K^0\gtrsim Nl$.\cite{simon03} For thick wires, such a condition is difficult to meet experimentally and the renormalization and no renormalization of the Kondo temperature occurs when we vary the wire gate voltage. To see the effects discussed in this paper,
it is therefore preferable to work with thin wires containing a few channels only.

We have also assumed that the wire is in a ballistic regime, or at least that the mean free path $l_d$ is the largest length scale by far. 
If the  mean free path $l_d$ would be for example smaller than
the length of the short wire, we would then need to compare the Kondo length scale
$\xi_K^0$ with $l_d$ and not with $2l$ the wire length. Moreover, it is very likely that the simple dependences of $T_K$ and of the RKKY interaction  with $E_W$ will be washed out. Nevertheless, for a clean semiconducting wire of typical length $\sim 1$ micron, we may expect $l_d> 2l$.   

We have also neglected  Coulomb interactions in the wire. Coulomb interactions will first affect both 
the Kondo temperature and the RKKY interactions. Assuming that the interacting wire can be described by a Luttiger liquid characterized by a dimensionless strength parameter $g$ ($g=1$ corresponds to the non interacting case), the Kondo temperature takes a power-law fashion 
$T_K^0\sim v_F (\rho_0 J)^{2/(1-g)}$ instead
of the conventional exponential dependence.\cite{lee-toner} It can also be shown that the RKKY interaction decreases as a power law, $I\sim v_F(\rho_0 J)^2/(2l)^g$, with  $2l$ the inter-impurity distance.\cite{egger}
For a Luttiger liquid with repulsive interactions $g<1$, it implies an increase of the Kondo temperature
(therefore a decrease of the screening cloud) and the RKKY interaction becomes longer range compared
to the non-interacting case. We have seen in the introduction for the non interacting case that 
the criterium for  FSE to occur is $2l \lesssim \xi_K^0$ whereas the criterium
$I> T_K^0$ provides the more restrictive condition  $2l \lesssim (\rho_0J)^2\xi_K^0$.
On the other hand, in presence of interactions, both criteria provides the same length scale
$l^*\sim (\rho_0 J)^{2/(g-1)}=\xi_K^0$ under which FSE and RKKY interaction come into play. Therefore, we expect here an even  sharper cross-over from a $1-$impurity regime
to a $2-$ impurity regime. Nevertheless, repulsive interaction considerably affects also
transport properties through the dots. In the low energy limit, processes that transfer one electron between the leads and the interacting wire become irrelevant.\cite{kim} Therefore the leads and the wire decouple as two independent(very anisotropic) channels. Because the coupling between the dot and the wire is much stronger, 
the screening of the impurity occurs at low energy in the wire only. The conductance in the leads is  dominated in this low energy limit by potential scattering terms that lead to resonant cotunneling processes (that do not renormalize).  The conductance is therefore very small compared to $G_U$ whatever a single impurity or a $2-$ impurity scenario occurs. It seems therefore less obvious  
to differentiate these two cases (and therefore to provide a lower estimate of $\xi_K^0$) in presence of interactions at least in the equilibrium conductance. The differential conductance may reveal on the other hand
interesting features at large bias that may help to distinguish the $1-$impurity regime from
the $2-$ impurity regime. A full quantitative analysis of transport properties incorporating consistently both Coulomb
interactions and finite level spacing is required but goes beyond the scope of the present paper.

\acknowledgments
We would like to acknowledge interesting conversations with I. Affleck, K. Le Hur and R. Lopez.



\begin{thebibliography}{999}
\bibitem{dot} D. Goldhaber-Gordon, H. Shtrikman, D. Mahalu, 
D. Abusch-Magder, U. Meirav, and M.A. Kaster, Nature {\bf 391}, 156 (1998).
\bibitem{Cronenwett} S.M. Cronenwett, T.H. Oosterkamp, and L.P. Kouwenhoven, 
Science {\bf 281}, 540 (1998); F. Simmel, R.H. Blick, U.P. Kotthaus,
W. Wegsheider, and M. Blichler, Phys. Rev. Lett. {\bf 83}, 804 (1999).
\bibitem{Wiel}
W.G. van der Wiel, S. De Franceschi, T. Fujisawa, 
J.M. Elzerman, S. Tarucha, and L.P. Kouwenhoven, Science, {\bf 289}, 2105
(2000).
\bibitem{affleck01} I. Affleck and P. Simon, Phys. Rev. Lett. {\bf 86}, 2854
  (2001); P. Simon and I. Affleck, Phys. Rev. {\bf B64}, 085308 (2001).
\bibitem{Hu} H. Hu, G.-M. Zhang, and Yu Lu, Phys. Rev. Lett. {\bf 86}, 5558 (2001).
\bibitem{sorensen04} E. S. Sorensen, I. Affleck, cond-mat/0409034 (2004). 
\bibitem{simon02} P. Simon and I. Affleck, Phys. Rev. Lett. {\bf 89} 206602 (2002).
\bibitem{simon03}  P. Simon and I. Affleck, Phys. Rev. {\bf B68}, 115304 (2003).
\bibitem{balseiro}  P. S. Cornaglia and C. A. Balseiro, Phys. Rev. Lett. {\bf 90}, 216801 (2003) 
\bibitem{thimm} W. B. Thimm, J. Kroha, J. von Delft, Phys. Rev. Lett. {\bf 82} 2143 (1999).
\bibitem{balseiro02} P. S. Cornaglia, C. A. Balseiro, Phys. Rev. B {\bf 66}, 115303 (2002);
ibid {\bf 66}, 174404 (2002).
\bibitem{barzykin} V. Barzykin and I. Affleck, Phys. Rev. {\bf B 61}, 6170 (2000).
\bibitem{doniach} S. Doniach, Physica {\bf 91B}, 231 (1977); S. Doniach, in Proceedings of the Conference on Itinerant Electron Magnetism, Oxford, England, 1976 (unpublished).
\bibitem{craig}  N. J. Craig, J. M. Taylor, E. A. Lester, C. M. Marcus, M. P. Hanson, and A. C. Gossard, Science {\bf 304}, 565 (2004).
\bibitem{galkin} S. Galkin, C. A. Balseiro and M. Avignon, Euro. Phys. J. B {\bf 38}, 519 (2004).
\bibitem{sorensen96} E. S. Sorensen and I. Affleck, Phys. Rev. {\bf B 53}, 9153-9167 (1996)
\bibitem{jones89} B. A. Jones, B. G. Kotliar, and A. J. Millis, Phys. Rev. {\bf B 39}, 3415 (1989). 
\bibitem{correa} A. Correa, K. Hallberg, and C. A. Balseiro, Europhys. Lett. {\bf 58}, 899 (2002).
\bibitem{jayaprakash} C. Jayaprakash, H. R. Krishna-murthy, and J. W. Wilkins, Phys. Rev. lett
{\bf 47}, 737 (1981).
\bibitem{sun05} P. Sun and G. Kotliar, cond-mat/0501176.
\bibitem{jones}
B. A.~Jones, C. M.~Varma, and J. W.~Wilkins, Phys. Rev. Lett. \textbf{58}, 843
  (1987); {\it ibid.}
\textbf{61}, 125 (1988). B. A. Jones and C. M. Varma, Phys. Rev.  \textbf{B 40} (1989).
\bibitem{affleck}
I.~Affleck and A. W .W. Ludwig, Phys. Rev. Lett. \textbf{68}, 1046 (1992). 
\bibitem{affleck95} I.~Affleck, A. W. W.~Ludwig, and B. A.~Jones, Phys. Rev. B\textbf{52}, 9528
  (1995).
\bibitem{schiller} Q.-f. Sun and H. Guo
Phys. Rev. {\bf B 64}, 153306 (2001), E. Lebanon and A. Schiller, Phys. Rev. {\bf B 65}, 035308 (2002).
\bibitem{ng} T. K. Ng and P. A. Lee, Phys. Rev. Lett. {\bf 61}, 1768 (1988).
\bibitem{mckenzie} S. Y. Cho, H.-Q. Zhou, and R. H. McKenzie, Phys. Rev. {\bf B 68}, 125327 (2003).
\bibitem{vavilov} M. G. Vavilov, L. I. Glazman, cond-mat/0404366.
\bibitem{simon04}  P. Simon, R. Lopez, and Y. Oreg, Phys. Rev. Lett. in press (cond-mat/0404540). 
\bibitem{review} L. I. Glazman and M. Pustilnik, J. Phys. Condens. Matter, {\bf 16}, R513 (2004).
\bibitem{lee-toner} D.-H. Lee and J. Toner, Phys. Rev. Lett. {\bf 69}, 3378 (1992); A. Furusaki and N. Nagaosa,  Phys. Rev. Lett. {\bf 72}, 892 (1994).
\bibitem{egger} R. Egger and H. Schoeller, Phys. Rev. B {\bf  54}, 16337 (1996); K. Hallberg and R. Egger, Phys. Rev. B {\bf 55}, R8646-R8649 (1997). 
\bibitem{kim} E. H. Kim, cond-mat/0106575 (unpublished).
\end{thebibliography}
\end{document}